# FFT-based solution of 2D and 3D magnetization problems in type-II superconductivity

**Leonid Prigozhin[1] and Vladimir Sokolovsky[2]**

[1]J. Blaustein Institutes for Desert Research, Ben-Gurion University of the Negev, Sede Boqer Campus  84990, Israel
[2]Physics Department, Ben-Gurion University of the Negev, Beer-Sheva 84105, Israel

E-mail: leonid@math.bgu.ac.il, sokolovv@bgu.ac.il

**Abstract**
We consider the Fast Fourier Transform (FFT) based numerical method for thin film magnetization problems [Vestgården and Johansen, SuST, 25 (2012) 104001], compare it with the finite element methods, and evaluate its accuracy. Proposed modifications of this method implementation ensure stable convergence of iterations and enhance its efficiency.

A new method, also based on the FFT, is developed for 3D bulk magnetization problems. This method is based on a magnetic field formulation, different from the popular *h*-formulation of eddy current problems typically employed with the edge finite elements. The method is simple, easy to implement, and can be used with a general current-voltage relation; its efficiency is illustrated by numerical simulations.

**1. Introduction**
Macroscopically, magnetization of type-II superconductors is well described by the eddy current model with a highly nonlinear current-voltage relation. Numerical simulations based on such a model help to understand the peculiarities of magnetic flux penetration into a superconductor and are necessary for the design of superconductor-based electronic and electrical devices. Thin superconducting films and bulk superconductors are two configurations of the main practical interest. A variety of finite element methods have been employed to solve numerically the thin film problems and also the bulk superconductor problems with, e.g., axial symmetry making them two-dimensional (2D).

A different approach to solving the thin film problems, the FFT-based approximation in space and the method of lines for integration in time, has been proposed in [1, 2]. This method was then employed by many authors, mainly for modeling thermal instabilities and flux avalanches in superconducting films (see [3-5] and the references therein). In this work we slightly modify the method to improve its efficiency and compare it with the finite element methods (Section 2).

Following the works [6-9], finite element methods for 3D bulk superconductor magnetization problems are now being intensively developed by several groups [10-15]. Recent numerical simulations for a cubic superconductor [12, 16-18] have demonstrated new interesting features of the current density distribution, absent in axially symmetric problems. To our knowledge, no FFT-based method has been proposed for 3D bulk magnetization problems. We develop such a method in Section 3 and compare it with the finite element methods using, in particular, simulation results for the benchmark problem 5 (magnetization of a cubic superconductor) in [19]. Our 3D FFT-based method uses the magnetic field as the main variable and is easy to implement: it employs a standard ordinary differential equations (ODE) solver for integration in time (the method of lines) and standard tools for the FFT and inverse FFT in space.

Although here we used only a field-independent isotropic power law current-voltage relation, the new 3D method is general and allows one to use a different relation to account for a critical current density dependence on the magnetic field, anisotropy of a superconductor, or flux cutting effects in the force free configurations. All numerical simulations in this work were performed in Matlab R2017a on a PC with Intel(R) Core(TM) i5-2400 CPU @ 3.10GHz, 16 GB RAM, Windows 7 (64-bit).

## 2. Thin film magnetization problems
*2.1 The 2D problem*

For completeness, we present briefly the problem formulation and numerical method [1]. Let, in the infinitely thin film approximation, a superconducting film occupy a 2D domain $\Omega$ in the plane $z=0$ and $h_{z,e}(t)$ be the normal to this plane component of the external magnetic field, assumed uniform for simplicity. The current, induced in the film by time variations of this field, satisfies

$$\nabla \cdot \boldsymbol{j} = 0 \quad \text{in} \quad \Omega, \quad j_n = 0 \quad \text{on} \quad \Gamma, \tag{1}$$

where $\boldsymbol{j}(x,y)$ is the sheet current density and $j_n$ is its normal component on the domain boundary $\Gamma$.

Let us assume at first that the film is simply connected (has no holes). Due to (1) there exists a stream function $g(x,y)$ such that

$$\boldsymbol{j} = \bar{\nabla} \times g \tag{2}$$

( i.e. $j_x = \partial_y g$, $j_y = -\partial_x g$ ) and $g = 0$ on $\Gamma$; we extend this function by zero to $\Omega_{\text{out}} = R^2 \setminus \Omega$.

In $\Omega$, we assume the nonlinear (power law) current-voltage relation,

$$\boldsymbol{e} = \rho(|\boldsymbol{j}|)\boldsymbol{j}, \tag{3}$$

where $\boldsymbol{e}$ is the electric field component tangential to the film, the sheet resistivity is $\rho(|\boldsymbol{j}|) = (e_0 / j_c)|\boldsymbol{j}/j_c|^{n-1}$, $e_0 = 10^{-4} \text{Vm}^{-1}$, the sheet critical current density $j_c$ and the power $n$ are assumed constant.

By the Faraday law

$$\mu_0 \dot{h}_z = -\nabla \times \boldsymbol{e}, \tag{4}$$

where $\mu_0$ is the magnetic permeability of vacuum, $\dot{f}$ means $\partial_t f$, and $\nabla \times \boldsymbol{e} = \partial_x e_y - \partial_y e_x$.

The Biot-Savart law yields

$$h_z = h_{z,e} + \nabla \times \int_\Omega G(\boldsymbol{r}-\boldsymbol{r}')\boldsymbol{j}(\boldsymbol{r}',t)\,\mathrm{d}\boldsymbol{r}', \tag{5}$$

where $\boldsymbol{r} = (x,y)$ and $G(\boldsymbol{r}) = (4\pi |\boldsymbol{r}|)^{-1}$ is the Green function; note the difference between the vectorial, $\bar{\nabla} \times g$, and scalar, $\nabla \times \boldsymbol{e}$, 2D curl operators. Since the sheet current density outside the film is zero, we can present the induced magnetic field in (5) via 2D convolutions: $h_z - h_{z,e} = (\partial_x G) * j_y - (\partial_y G) * j_x$. In terms of the stream function this equation becomes

$$h_z - h_{z,e} = -(\partial_x G)*(\partial_x g) - (\partial_y G)*(\partial_y g). \tag{6}$$

Taking Fourier transform and noting that the 2D Fourier transform (angular frequency, non-unitary, here and below) of $G$ yields $\tilde{G}(\boldsymbol{k}) = (2|\boldsymbol{k}|)^{-1}$ with $\boldsymbol{k} = (k_x, k_y)$, see [20], we arrive at $F(h_z - h_{z,e}) = \dfrac{|\boldsymbol{k}|}{2} F(g)$ or

$$g = F^{-1}\left(\frac{2}{|\boldsymbol{k}|} F\left[h_z - h_{z,e}\right]\right) - C, \tag{7}$$

where, to avoid ambiguity, we set $\frac{2}{|\mathbf{k}|}F(h_z - h_{z,e}) = 0$ for $|\mathbf{k}|=0$ and introduce a time-dependent constant $C(t)$ such that $\int_{\Omega_{\text{out}}} g\, d\mathbf{r} = 0$; note that $F(h_z - h_{z,e})|_{\mathbf{k}=0} = 0$ since $\int_{R^2}(h_z - h_{z,e}) d\mathbf{r} = 0$. Differentiating with respect to time we arrive at the differential equation for $g$,

$$\dot{g} = F^{-1}\left(\frac{2}{|\mathbf{k}|}F[\dot{h}_z - \dot{h}_{z,e}]\right) - \dot{C}, \tag{8}$$

in which the unknown shift $\dot{C}$ is determined by the integral condition $\int_{\Omega_{\text{out}}} \dot{g}\, d\mathbf{r} = 0$. Here $\dot{h}_z$ inside the film can be straightforwardly expressed via $g$: substituting (3) into (4) and taking (2) into account one obtains

$$\dot{h}_z = \frac{1}{\mu_0}\nabla\cdot\left[\rho(|\nabla g|)\nabla g\right]\quad \text{in } \Omega. \tag{9}$$

However, to compute $\dot{g}$ using the Fourier transform in (8) one needs to know $\dot{h}_z$ also in $\Omega_{\text{out}}$. To integrate the equation (8) in time for a given initial condition, it is needed to solve the following problem: for any time $t$, given $g$, find such $\dot{h}_z$ in $\Omega_{\text{out}}$ that $\dot{g}$ in (8) satisfies $\dot{g} = 0$ in $\Omega_{\text{out}}$. This problem can be solved iteratively as follows.

First, we find $\dot{h}_{z,\text{in}} = \dot{h}_z|_\Omega$ using (9) and define an initial approximation $\dot{h}^0_{z,\text{out}}$ in $\Omega_{\text{out}}$; if needed, a shift $\dot{h}^0_{z,\text{out}} := \dot{h}^0_{z,\text{out}} - A^0$, where $A^0$ is a constant, is applied to satisfy the condition

$$\int_{R^2}(\dot{h}_z - \dot{h}_{z,e})\,d\mathbf{r} = 0. \tag{10}$$

On the $i$-th iteration we first set, see (8),

$$\dot{g}^i = F^{-1}\left(\frac{2}{|\mathbf{k}|}F[\dot{h}^i_z - \dot{h}_{z,e}]\right) - C^i, \tag{11}$$

where the constant $C^i$ is such that $\int_{\Omega_{\text{out}}} \dot{g}^i d\mathbf{r} = 0$. Then compute

$$\dot{h}^{i+1}_{z,\text{out}} = \dot{h}^i_{z,\text{out}} - \lambda F^{-1}\left[\frac{|\mathbf{k}|}{2}F(\dot{g}^i_{\text{out}})\right]\bigg|_{\Omega_{out}} - A^i, \tag{12}$$

where the constant $A^i$ ensures (10) for $\dot{h}^{i+1}_z$, $\lambda > 0$ is a relaxation parameter (in [1, 2] $\lambda = 1$), and

$$\dot{g}^i_{\text{out}} = \begin{cases} \dot{g}^i & \text{in } \Omega_{\text{out}}, \\ 0 & \text{in } \Omega. \end{cases}$$

Provided the iterations converge, $\dot{g}^i_{\text{out}}$ converges to zero and the sought derivative $\dot{h}_z$ in $\Omega_{\text{out}}$ is found.

Finally, integrating the differential equation (8) for the stream function, one can use (2) to compute the sheet current density $\mathbf{j}$, the electric field $\mathbf{e}$ in the film is determined by (3), and the magnetic field

$$h_z = h_{z,e} + F^{-1}\left[\frac{|\mathbf{k}|}{2}F(g)\right].$$

*2.2 Numerical algorithm: implementation*

We use a uniform $N_x \times N_y$ grid in a rectangular domain $D = \{(x,y)\mid |x|\leq L_x,\ |y|\leq L_y\}$ containing the film domain $\Omega$, and replace the transforms $F, F^{-1}$ by the FFT and inverse FFT, respectively. Although the

derivation above holds only for the infinite plane, previous simulations showed that such an approximation can be sufficiently accurate even if the free space domain $\tilde{\Omega}_{\text{out}} = D \setminus \Omega$ is not large. The stream function $g$ is sought in the grid points; we employ the method of lines and use a standard Matlab ODE solver to integrate the system of equations $\dot{g} = f(g,t)$, where now $g$ denotes the $N_x \cdot N_y$-long vector of the grid values and the right hand side function $f$ realizes the discretized version of the iterative algorithm (11)-(12). Although our ODE system seems to be moderately stiff, we were unable to employ any Matlab ODE solver for stiff systems: these solvers use the Jacobian of $f$ (or its approximation), a too large non-sparse matrix in our case. The best results were obtained with the ode23 solver. It was ran with the default values of parameters (the relative tolerance $10^{-3}$, the absolute tolerance $10^{-6}$).

Randomly chosen one-sided finite differences were used in [1, 2] to approximate the derivatives in (9). Here we differentiated in the Fourier space and used the Gaussian filter to smooth the results. Thus, $\nabla g$ was approximated by $F^{-1}\left[i\boldsymbol{k}\exp(-|\boldsymbol{k}|^2 \sigma^2/2) F(g)\right]$, where $F$ is now the discrete Fourier transform. The parameter $\sigma$ should be of the order of the grid steps, $\Delta x$ and $\Delta y$; we chose $\sigma = \sqrt{\Delta x^2 + \Delta y^2}$ in our examples below. The same smoothing was used in (12). A different smoothing could be applied on a post-processing stage to compute the electric field; however, using a filter with a higher value of $\sigma$ to suppress scattering of $\boldsymbol{e} \sim |\boldsymbol{j}|^{n-1}\boldsymbol{j}$ did not improve the accuracy of $\boldsymbol{e}$ in our simulations.

The relaxation parameter $\lambda = 0.7$ was used in all simulations. Its introduction accelerated convergence of iterations and, usually, convergence was fast (the initial approximation $\dot{h}_{z,\text{out}}^0$ was taken as $\dot{h}_{z,\text{out}}$ from the previous time step). Sometimes, however, the iterations (11)-(12) failed to converge. In these seldom cases a very big fictitious $\dot{g}$ value in all grid points was returned by our function $f$ to the ODE solver. The solver, following its automatic time step control algorithm, decreased the integration time step, after which the iterations converged with the required tolerance.

It should be noted that (3) is valid only in $\Omega$ and the eddy current model does not determine the electric field outside the conducting media. On the other hand, the values of $\dot{h}_{z,\text{in}}^0$ in the nearest to the boundary $\Gamma$ grid points of $\Omega$, computed via (4) re-written as (9), may depend on the electric field in the grid points just outside the boundary. A simple but, possibly, crude way we used to overcome this complication was to define in $\tilde{\Omega}_{\text{out}}$ a sufficiently high resistivity $\rho_{\text{out}}$ and assume there the Ohm law, $\boldsymbol{e} = \rho_{\text{out}} \boldsymbol{j} = \rho_{\text{out}} \bar{\nabla} \times g$. This artificial current-voltage relation influences mainly the $\dot{h}_z$ values in the nearest to $\Gamma$ grid points through $\nabla \times \boldsymbol{e}$: the iterations (11)-(12), which do not change $\dot{h}_{z,\text{in}}$, efficiently eliminate the stray current in $\tilde{\Omega}_{\text{out}}$.

The stream function formulations of thin film problems can be extended to multiply connected films in several ways; see, e.g., [2, 21, 22]. In one of our examples below we simply filled in the hole and postulated there the Ohm law with a high enough constant resistivity to make the sheet current density in the holes negligibly small. We note that in this case (9) was applied in the domain $\Omega_{\text{filled}}$, with the power law current voltage relation in $\Omega$ and the linear Ohm law in the holes. The iterations (11)-(12) were performed in $\tilde{\Omega}_{\text{out}} = D \setminus \Omega_{\text{filled}}$.

*2.3 Thin film magnetization: numerical simulations*
In our simulations we used the dimensionless variables

$$(x', y') = \frac{(x, y)}{l}, \quad t' = \frac{t}{t_0}, \quad e' = \frac{e}{e_0}, \quad j' = \frac{j}{j_c}, \quad h'_z = \frac{h_z}{j_c}, \quad g' = \frac{g}{j_c l},$$

where $l$ is the characteristic film size and $t_0 = \mu_0 j_c l / e_0$. For simplicity, below in this section the prime is omitted. In all examples the initial sheet current density was zero, and $h_{z,e} = t$. The iterations (11)-(12) were performed until the mean value of $|g^i|$ in $\tilde{\Omega}_{\text{out}}$ becomes less than $2 \cdot 10^{-4}$.

Our first example is the thin disk magnetization problem having an analytical solution $h_z, j$ for the Bean model [23, 24]. As in [25], the corresponding electric field $e(r,t)$ was calculated numerically integrating the Faraday law $r^{-1} \partial_r (re) = -\partial_t h_z$ with $e|_{r=0} = 0$. The power law model converges to the Bean model as the power $n$ tends to infinity [26], and we compared with this analytical solution our simulation results for $n = 1000$ and the disk of radius one in the square $D = \{(x, y) \mid |x| \leq 2, |y| \leq 2\}$. Presented results (see Table 1 and Figure 1) are for $h_{z,e} = 0.5$, $\rho_{\text{out}} = 1$, and different grids. The relative deviations from the analytical solution (in the $L^1$ norm) of the current density and normal to the film magnetic field component, $\delta j$ and $\delta h_z$, respectively, for the $512 \times 512$ grid were similar to those obtained using the finite element approximation of the electric field problem formulation in [25] and several times higher than those for the mixed finite element formulation in [21] obtained using a mesh of 4200 finite elements in the disk. The finite element methods were faster in this example.

We note that in the infinitely thin film approximation the magnetic field on the film boundary is singular and this explains why the field error $\delta h_z$ is typically higher than the current density error $\delta j$. Similar $\delta j$ deviations for the two finest grids in Table 1 mean, probably, that the remaining deviation is related not to the mesh resolution but to replacement of the infinite plane by a finite domain, convergence of iterations with the given tolerance, the finite value of $n$, etc.

Small variations of current density near the critical current density value lead to large fluctuations of the electric field if the power $n$ is high; in this example we could not obtain any accurate approximation of the electric field using the FFT based method (see Figure 1). The methods based upon dual and mixed variational formulations, [25] and [21], respectively, were able to compute the electric field with deviations $\delta e$ equal to 4.3% and 1.1%, respectively. However, the FFT based method is much simpler and easier to implement, which is an advantage. Fast growth of computation time with the number of grid points (see Table 1) suggests that the ODE system becomes stiffer and the ODE solver uses smaller time steps for finer grids.

**Table 1**. Thin disk. Deviation of the numerical $(n = 1000)$ solution from the Bean model.

| $N_x, N_y$ | CPU time, min | $\delta j, \%$ | $\delta h_z, \%$ |
|---|---|---|---|
| 128 | 0.2 | 2.2 | 12 |
| 256 | 2.0 | 1.0 | 7.1 |
| 512 | 49 | 0.9 | 4.0 |

Another analytical Bean model solution is known for an infinite strip [27]. We solved numerically the magnetization problem for a rectangular film, $\Omega = \{(x, y) \mid |x| \leq 3.2, |y| \leq 0.5\}$, using the $1024 \times 256$ mesh in $D = \{(x, y) \mid |x| \leq 4, |y| \leq 1\}$. All other parameters were as in the previous example. Away from the film ends $x = \pm 3.2$ the solution is close to that for the infinite strip: see Figure 2, where the Bean model solution is compared to the numerical one for all grid points with $|x| < 2$. Now the calculated electric field is surprisingly accurate, possibly, because the grid is well aligned with the boundary of $\Omega$.

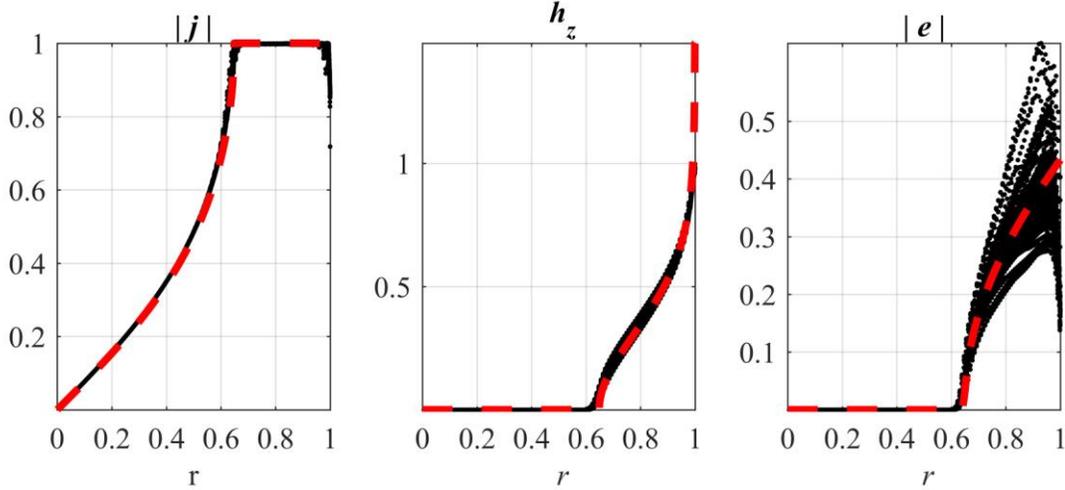

**Figure 1.** Thin disk, $h_{z,e} = 0.5$. Shown: $|j|, h_z, |e|$ as functions of radius $r$. The Bean model solution (dashed red line) and numerical solution (black dots, for all grid points in the disk) for $n = 1000$ and the $512 \times 512$ mesh.

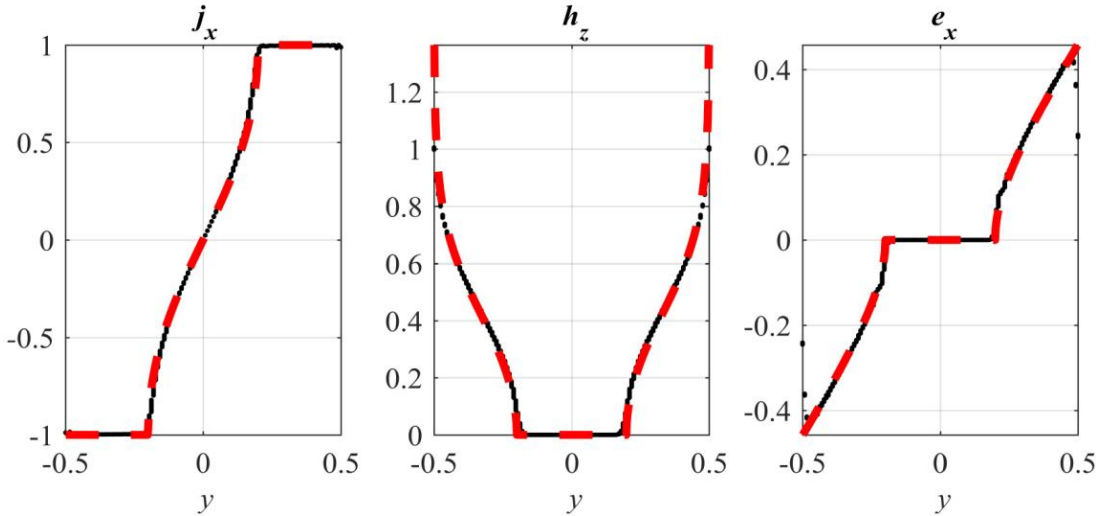

**Figure 2.** Rectangular film, $h_{z,e} = 0.5$. Shown: $j_x, h_z, e_x$ as functions of $y$. The Bean model solution for an infinite strip (dashed red line) and numerical solution (black dots, the values in the film grid points away from the film ends) for $n = 1000$ and the $1024 \times 256$ mesh.

For our next example (Figure 3) we chose a rectangular film $\Omega_{\text{filled}} = \{(x, y) \mid |x| \leq 1.5, |y| \leq 1\}$ with a circular hole $(x - 0.5)^2 + (y - 0.25)^2 \leq 0.5^2$, set $n = 50$, and used the $768 \times 512$ mesh in $D = \{(x, y) \mid |x| \leq 2.25, |y| \leq 1.5\}$. The growing magnetic field penetrates the film from its boundary and, at some moment in time, there appears a narrow runway through which a significant part of magnetic flux reaches the circular hole; the electric field is especially strong along this runway. To suppress the stray current in the outer space near the ends of this runway we set $\rho_{\text{out}} = 500$ both in the hole and outside the film. It took 63 minutes to compute the solution at $t = 0.4$ by

the FFT method; this solution (Figure 3, left) can be compared to the one obtained by the mixed finite element method [21] (Figure 3, right). In the latter case a finite element mesh in $\Omega_{\text{filled}}$ contained 8 thousand elements, the power law with $j_{c,\text{hole}} = 0.02 j_c$ was set in the filled hole, and the computation time was 87 minutes. Numerical smearing of the infinitely thin runway from the outer film boundary to the hole is slightly more pronounced for the FFT method; probably, this is a result of smoothing. Significant fluctuations of the electric field computed by the FFT-method can be noticed in the penetration zone surrounding the circular hole.

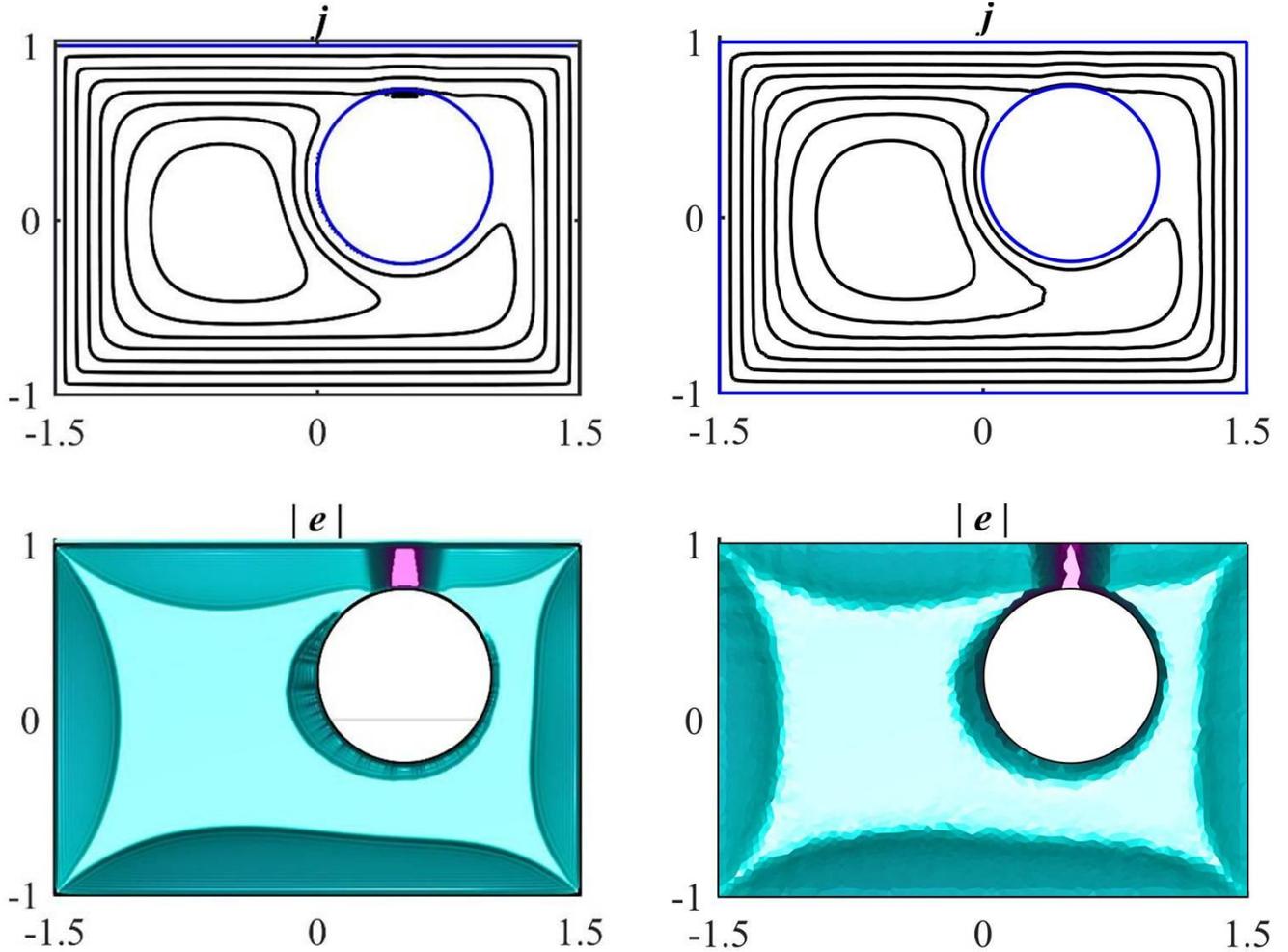

**Figure 3.** Film with a circular hole, $h_{z,e} = t$, $n = 50$. Numerical simulation results for $t = 0.4$ obtained using the FFT-based method (left) and by the finite element method [21] (right). Shown: the current lines (up) and $|e|$ levels (bottom).

## 3. Bulk magnetization problems

An FFT based method for thin film problems can be derived also with the magnetic field component $h_z$ as the main variable, not the stream function. Without going into details we note that we explored such an approach and found it less accurate, probably, because $h_z$ is singular on the film boundary. Hence, using the stream function, continuously extendable by zero outside the film, is preferable. The 3D analogue of the 2D stream function $g$ is the vector current potential $T$ such that $j = \nabla \times T$. Contrary to the 2D case, in 3D problems only tangential component of $T$ can be made zero on the boundary of a superconductor, a gauge should be chosen and implemented, a scalar magnetic potential is usually needed as well. Although development of a numerical method combining the vector current potential and FFT deserves special investigation, here we present a very simple 3D numerical method derived for the magnetic field $h$. Unlike the thin film case, on the boundary of a bulk superconductor magnetic field remains finite and even continuous (the first critical field is neglected).

Our formulation is different from the popular $h$-formulation of the eddy current problems implemented with the edge finite elements. Our method requires less empty space around the superconductor in the computational domain.

*3.1 Bulk problems: formulation and numerical method*

Let now the superconductor occupy a 3D domain $\Omega$ with the boundary $\Gamma$, $h_e(t)$ be the external magnetic field, for simplicity assumed uniform, and $j$ denote the superconducting current density satisfying

$$\nabla \cdot j = 0 \quad \text{in} \quad \Omega, \quad j_n = 0 \quad \text{on} \quad \Gamma.$$

We assume the power law current-voltage relation,

$$e = \rho(|j|)j, \tag{13}$$

where $e$ is the electric field, $\rho(|j|) = (e_0/j_c)|j/j_c|^{n-1}$, the power $n$ and the critical current density $j_c$ are constants.

To model magnetization of this superconductor, let us consider the following problem. Given magnetic field $h$, the current density is determined by the Ampere law, $j = \nabla \times h$, and we can find $e$ in $\Omega$ using (13). By the Faraday law

$$\mu_0 \dot{h} = -\nabla \times e, \tag{14}$$

so the time derivative of $h$ is known in $\Omega$. We assume $h$ is such that $j = 0$ in $\Omega_{\text{out}} = R^3 \setminus \Omega$ and want to find $\dot{h}|_{\Omega_{\text{out}}}$ for which the time derivative of the Biot-Savart law,

$$\dot{h} = \dot{h}_e + \Phi[\dot{j}] = \dot{h}_e + \Phi[\nabla \times \dot{h}], \tag{15}$$

holds with $\dot{j}|_{\Omega_{\text{out}}} = \nabla \times \dot{h}|_{\Omega_{\text{out}}} = 0$. Here $\Phi[j] = \nabla \times \int_{\Omega} G(r - r')j(r', t)\,dr'$, where the Green function is $G(r) = (4\pi |r|)^{-1}$ as above but now $r = (x, y, z)$.

The operator $\Phi$ can be presented via convolutions in $R^3$,

$$\Phi[j] = \begin{cases} j_z * \partial_y G - j_y * \partial_z G \\ j_x * \partial_z G - j_z * \partial_x G \\ j_y * \partial_x G - j_x * \partial_y G \end{cases}. \tag{16}$$

Since the Fourier transform of $G$ in $R^3$ is $\tilde{G}(k) = 1/|k|^2$, where $k = (k_x, k_y, k_z)$, see [20], (16) can be written as

$$\Phi[j] = F^{-1}\left\{\frac{i}{|k|^2}\begin{bmatrix} k_y F(j_z) - k_z F(j_y) \\ k_z F(j_x) - k_x F(j_z) \\ k_x F(j_y) - k_y F(j_x) \end{bmatrix}\right\}. \tag{17}$$

Taking into account that $\int_{R^3}(h - h_e)\,dr$ should be zero at each moment in time, for $k = 0$ we replace $1/|k|^2$ by zero. Clearly, $\nabla \times \Phi[j] = \nabla \times (h - h_e(t)) = j$.

Our iterative solution $\dot{\boldsymbol{h}}$ of the problem above is as follows. Let at time $t$ the magnetic field $\boldsymbol{h}$ be known. Find $\boldsymbol{j} = \nabla \times \boldsymbol{h}$, compute

$$\boldsymbol{e} = \begin{cases} \rho(\boldsymbol{j})\boldsymbol{j} & \text{in } \Omega \\ \rho_{out}\boldsymbol{j} & \text{in } \Omega_{out} \end{cases}$$

where $\rho_{out}$ is the fictitious resistivity, chosen as in the 2D problem to define the electric field and suppress the stray current in the closest neighborhood of $\Omega$. Then set $\dot{\boldsymbol{h}}_{in} = -\mu_0^{-1}\nabla \times \boldsymbol{e}|_{\Omega}$ and define an initial approximation, $\dot{\boldsymbol{h}}_{out}^0$, in $\Omega_{out}$. On the $i$-th iteration, compute $\dot{\boldsymbol{j}}^i = \nabla \times \dot{\boldsymbol{h}}^i$ and set

$$\dot{\boldsymbol{h}}_{out}^{i+1} = \dot{\boldsymbol{h}}_e + \Phi[\dot{\boldsymbol{j}}_{in}^i]|_{\Omega_{out}}, \tag{18}$$

where $\dot{\boldsymbol{j}}_{in}^i = \dot{\boldsymbol{j}}^i$ in $\Omega$ and zero in $\Omega_{out}$. Provided these iterations converge, $\nabla \times \dot{\boldsymbol{h}}|_{\Omega_{out}} = (\dot{\boldsymbol{j}}_{in})|_{\Omega_{out}} = \boldsymbol{0}$ as desired.

Such iterative algorithm works well if $\Omega_{out}$ is a contourwise simply connected domain, i.e. every closed contour in $\Omega_{out}$ can be presented as a boundary of a surface also belonging to $\Omega_{out}$. The case of a multiply connected domain is more complicated. Indeed, as a result of such iterations we arrive at a field $\dot{\boldsymbol{h}} = \dot{\boldsymbol{h}}_e + \Phi[\dot{\boldsymbol{j}}]$ satisfying $\nabla \cdot \dot{\boldsymbol{h}} = 0$ and $\nabla \times \dot{\boldsymbol{h}} = 0$ in $\Omega_{out}$ and having a continuous normal component on $\Gamma$ (since $\nabla \cdot \dot{\boldsymbol{h}} = 0$ in all $R^3$). Furthermore, this normal component, $\dot{h}_n$, is given because we know $\dot{\boldsymbol{h}}$ in $\Omega$. If $\Omega_{out}$ is a contourwise simply connected domain, such a field is unique. For a multiply connected domain it is not, since the problem

$$\nabla \cdot \dot{\boldsymbol{h}} = 0, \ \nabla \times \dot{\boldsymbol{h}} = 0 \text{ in } \Omega_{out}, \ \dot{h}_n = 0 \text{ on } \Gamma$$

has nontrivial solutions and additional conditions are necessary [28, 29]. To model magnetization of a bulk superconductor topologically equivalent to torus (our last example below), we circumvented this difficulty cutting the hole by a layer of a normal conductor with high resistivity. This makes the conducting and non-conducting domains simply connected, while the current in this artificial layer remains negligibly small.

Also in many other aspects the implementation of our algorithm is similar to that for the thin film problems. We define a uniform $N_x \times N_y \times N_z$ grid in the computational domain $D = \{(x,y,z)|\ |x| \leq L_x, |y| \leq L_y, |z| \leq L_z\}$ containing $\Omega$ and some empty space around it, replace the Fourier transform in (17) by FFT, employ FFT with the Gaussian smoothing for computing all necessary derivatives, and use the method of lines (with the Matlab ODE solver ode23) to integrate in time the equations $\dot{\boldsymbol{h}} = f(t, \boldsymbol{h})$, where now $\boldsymbol{h}$ denotes the array of grid node vector values of the magnetic field. The iterations (18) were stopped when the average of the node values of $|\dot{\boldsymbol{h}}^{i+1} - \dot{\boldsymbol{h}}^i|$ in $\Omega_{out}$ is less than $\delta_{it} \max(|\dot{\boldsymbol{h}}_e|, 1)$, where $\delta_{it}$ is the prescribed tolerance.

*3.2 Bulk magnetization: simulation results*

We used dimensionless variables,

$$(x', y', z') = \frac{(x,y,z)}{l}, \quad t' = \frac{t}{t_0}, \quad e' = \frac{e}{e_0}, \quad j' = \frac{j}{j_c}, \quad h' = \frac{h}{j_c l}, \tag{19}$$

where $l$ is the characteristic size and $t_0 = \mu_0 j_c l^2 / e_0$; below, the prime will be omitted.

Penetration of the magnetic field $\boldsymbol{h}_e(t) = (0,0,t)$ into a hollow ball $\Omega = \{0.5 \leq |r| \leq 1\}$ was one of the axisymmetric 2D problems solved in [30]. We now solved this problem without using the symmetry (Figure 4).

We assumed $n=30$, used the Gaussian smoothing with $\sigma=d$, where $d=\sqrt{\Delta x^2+\Delta y^2+\Delta z^2}$, and set $\rho_{out}=5$, $\delta_{it}=10^{-4}$. For the $128\times 128\times 128$ mesh in the domain $D=\{(x,y,z)|\,|\mathrm{x}|\leq 1.5,|y|\leq 1.5,|z|\leq 1.5\}$ solution of this problem for $0\leq t\leq 0.4$ took 2 hours; the solution obtained is similar to that in [30]. While the external field is low, the hole plays no role and the field penetration depth for the Bean model can be found asymptotically (see [31]). For $t=0.2$ the calculated depth agrees very well with this asymptotic solution (Figure 4, top). The hollow ball is contourwise simply connected and solution is unique also after magnetic field starts to penetrate into the hole.

Our next example is magnetization of a cubic superconductor. Choosing the characteristic size $l$ equal to half the cube edge and rescaling the variables according to (19), for the benchmark problem 5 from [19] in the dimensionless variables one gets: $\Omega=\{|x|\leq 1,|y|\leq 1,|z|\leq 1\}$ and $\mathbf{h}_e=(0,0,A\sin(\omega t))$ with $A=0.3191, \omega=9869$. As in [19], we set $n=100$ and solved the problem for $0\leq t\leq T=\pi/2\omega$ ($T$ is the first quarter of the period, the initial magnetization) using the same grid and domain $D$ as in the previous example. The external field changes now much faster. Hence, the electric field is stronger and, to suppress the stray current in the close-to-boundary outside layer, we used $\rho_{out}=5000$. Results of modeling the magnetic field penetration are presented in Figure 5.

Appearance of a significant $z$ component of $\mathbf{j}$, necessary for shielding the zero-field core, is an unexpected feature of current density distribution discovered and explained in [16-18]. We found that smoothing, employed in the FFT-based method, smears the narrow peaks of $j_z$ and, therefore, had to use a weaker Gaussian filter with $\sigma=0.5d$ to arrive at the $j_z$-distribution (Figure 6) similar to that in [19]. The maximal value, $\max j_z=0.273$, was achieved in the cross-sections $z=\pm 0.773$; one of the points of the maximum was $(x_0,y_0,z_0)=(0.469,-0.516,-0.773)$. These values are close to those in [19] (cf. Figure 7 in [19] and our Figure 6, right). It should be noted that increasing $\rho_{out}$ and decreasing $\sigma$ make the problem stiffer: in this example the computation time was about 15 hours.

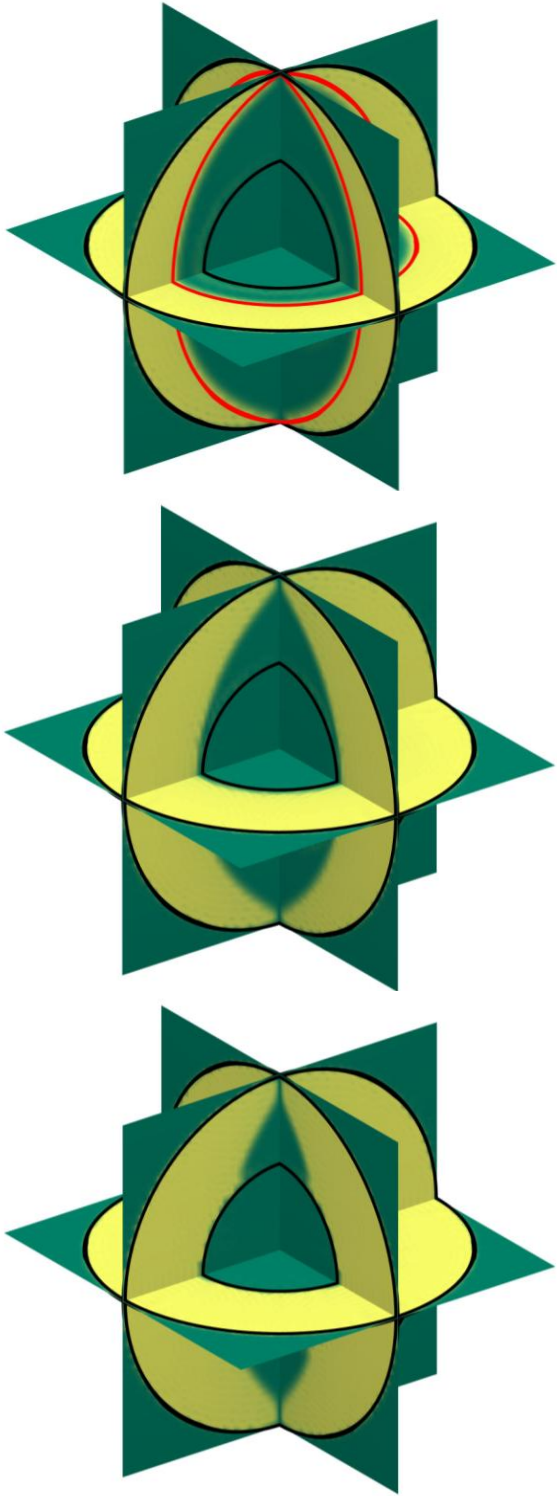
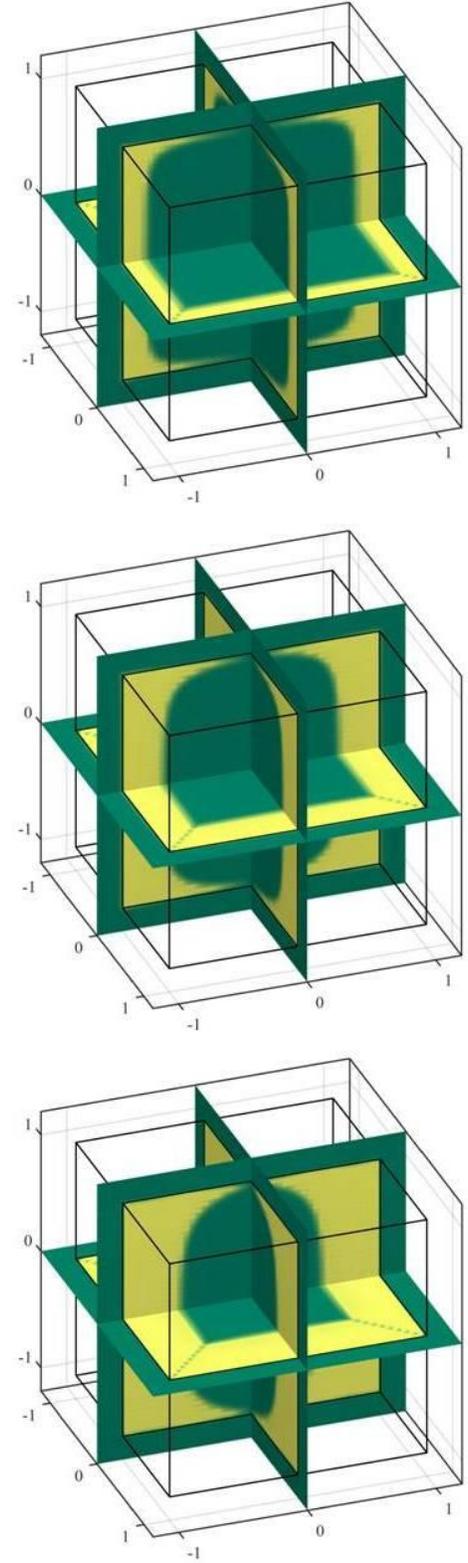

**Figure 4.** Magnetic field $h_{e,z}=t$ penetrates a hollow ball. Shown: $|j|$ levels for $t=0.2$ (top), $0.3$ (middle), $0.4$ (bottom). The red line indicates the asymptotic penetration depth.

**Figure 5.** Growing field $\boldsymbol{h}_e=(0,0,h_{e,z}(t))$ penetrates into a cubic superconductor; $|j|$ levels for $t=0.25T, 0.5T$ and $T$ (from top to bottom).

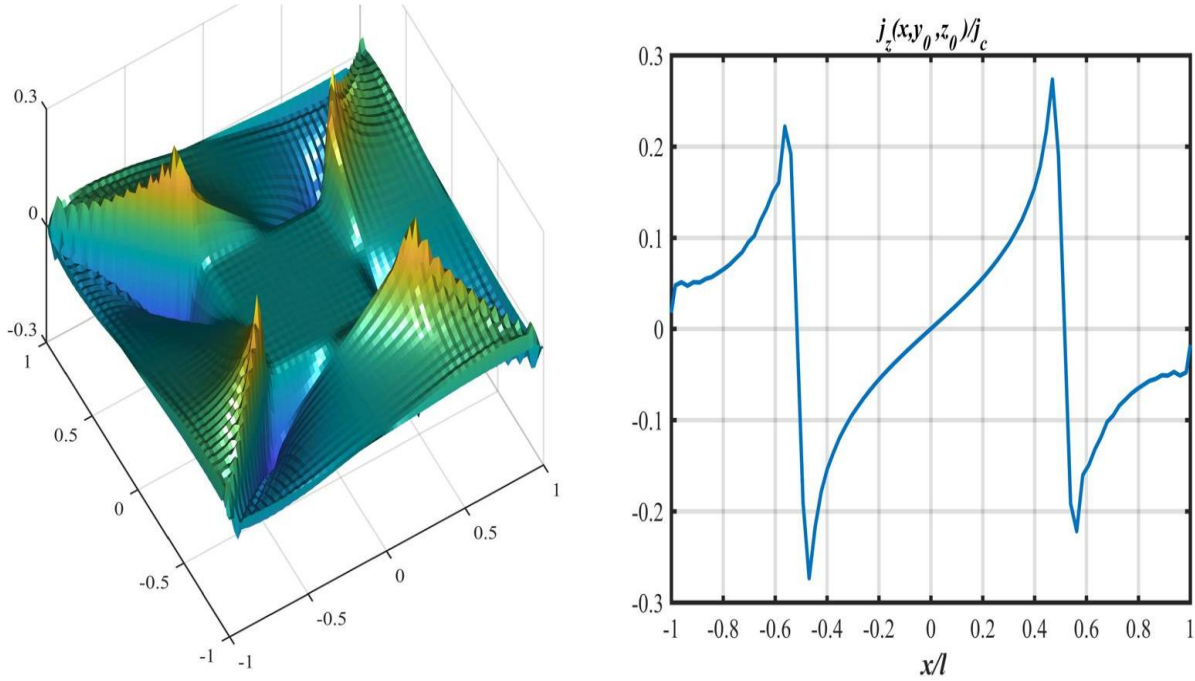

Figure 6. Left: $j_z$ in the cross-section $z_0 = -0.773$, where the maximal value, $\max j_z = 0.273$, is achieved. Right: $j_z(x, y_0, z_0)$ for $y_0 = -0.516$ and the same $z_0$.

Our last example is magnetization of a rectangular prism $\{(x,y,z) |\ |x| \leq 1.5, |y| \leq 1, |z| \leq 0.5\}$ having a cylindrical hole $\{(x,y,z) | (x-0.5)^2 + (y-0.25)^2 < 0.5^2\}$; all $z = const$ cross-sections of this domain are exactly the 2D domain in Figure 3. To make the conducting and non-conducting domains simply connected, we introduced a cutting-hole layer $\Omega_0 = \{(x,y,z) | (x-0.5)^2 + (y-0.25)^2 \leq 0.5^2,\ |z| \leq 0.05\}$ of a normal material having the same resistivity $\rho_{out}$ as is assumed for the outer space $\Omega_{out} = R^3 \setminus (\Omega \cup \Omega_0)$. The iterations (18), intended to eliminate the stray outside current, were applied in $\Omega_{out}$ but not in $\Omega_0$. For $\boldsymbol{h}_e(t) = (0,0,t)$ the resistivity $\rho_{out} = 200$ was sufficient to make the current in $\Omega_0$ negligibly small even without such iterations (see the hole $|\boldsymbol{j}|$ levels in Figure 7, left). In this example we set $n = 50, \sigma = d, \delta_{it} = 2 \cdot 10^{-4}$, and used the $192 \times 128 \times 64$ mesh in the box with the sides 1.5 times the prism sides, $D = \{(x,y,z) |\ |x| \leq 2.25, |y| \leq 1.5, |z| \leq 0.75\}$. Computing solution for $0 \leq t \leq 0.3$ took 4.5 hours. As in the thin film case, the electric field becomes very strong along a narrow layer connecting the hole with the outside space. Although strongly smeared in numerical solution, this phenomenon is clearly seen in Figure 7, right (we plotted the levels of $\min(|e|,1)$ to make distinguishable also other features of the electric field distribution). Our Matlab program, used to compute this example and plot the results, can be downloaded from [32]. Adapting this program to other geometries and/or current-voltage relations should not be difficult (see the Readme.pdf file in [32])

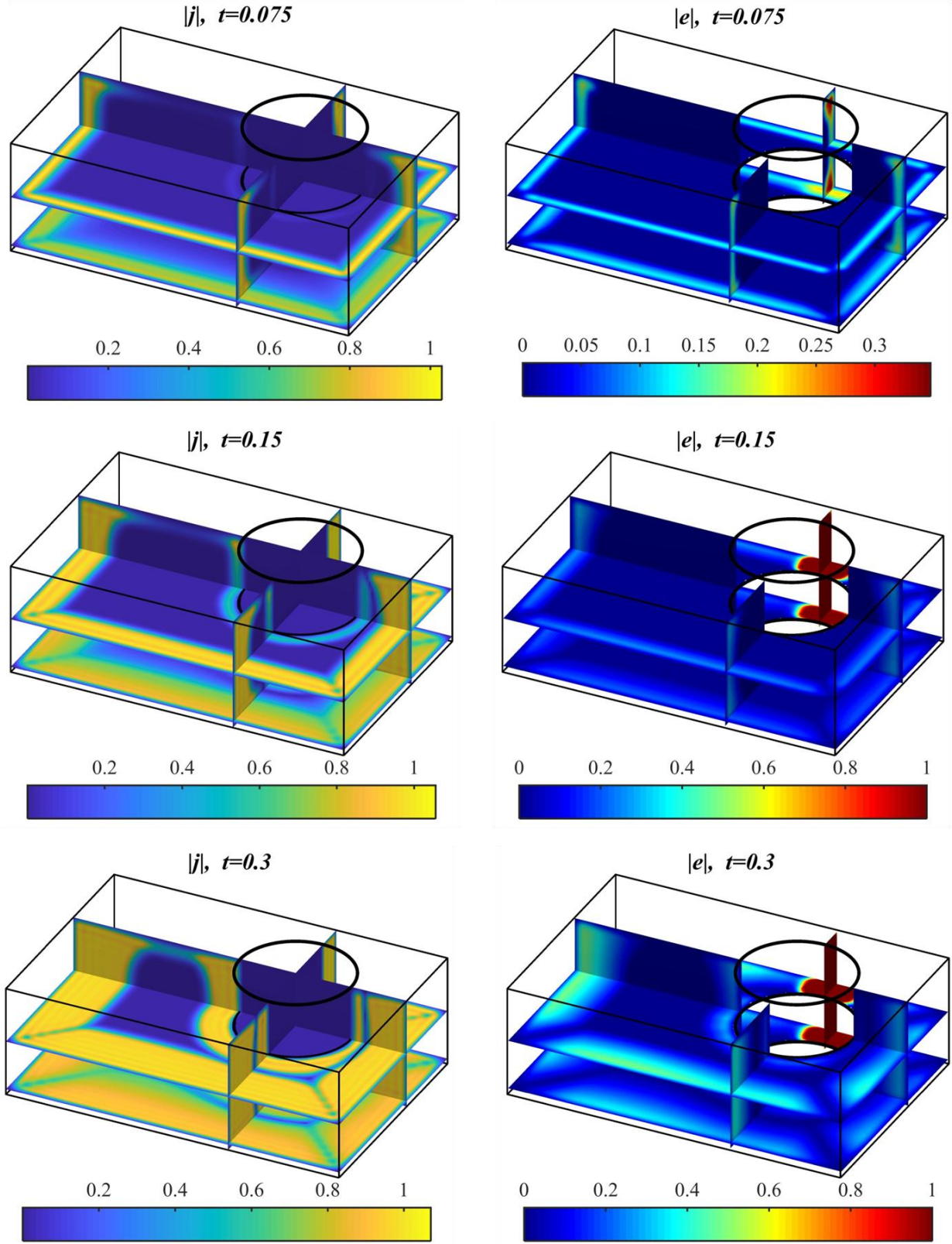

Figure 7. Magnetization of a rectangular prism with a hole: $|\boldsymbol{j}|$ levels (left) and $\min(|\boldsymbol{e}|,1)$ levels (right) in the cross-sections $z=-0.45, z=0, x=0.5,$ and $y=0.25$.

**Conclusion**

The FFT-based method [1, 2] for thin film magnetization problems continues to gain popularity and presents an alternative to the finite element methods. Our work has started as an attempt to compare the accuracy and efficiency of these two approaches. We introduced several modifications into the original implementation of the FFT-based method. Thus, using an under-relaxation in iterations (11)-(12), employing an ODE solver with automatic time step control, and using this control also to ensure convergence of these iterations, we stabilized the numerical solution and made it more efficient. We also replaced the randomly alternating forward and backward one-sided finite differences, used to approximate the spatial derivatives in [2], by a regular deterministic procedure: differentiation in the Fourier space. Using several test problems we showed that, typically, the FFT-based method is accurate and efficient, even though it can be difficult to reach the accuracy of some finite element methods. The FFT-method parameters, $\sigma$ and $\rho_{\text{out}}$, influence stiffness of the ODE system and their choice may need a compromise between the accuracy and efficiency.

Computing electric field for the power law current-voltage relation can be difficult if the power is high; this complication is well known and we found surprising that, at least if the domain boundary is well aligned with the rectangular grid, the FFT method could overcome this difficulty rather well. Although the mixed and dual finite element methods [21, 25], especially derived to accurately approximate the electric field in thin films, outperform the FFT method in this respect, their implementation is significantly more complicated.

Simplicity is a very attractive property of the FFT-based method for thin film problems. In addition, using the method of lines for integration in time makes it very easy to replace the isotropic power law by a different current-voltage relation.

The main result of our work is, however, development of a new 3D FFT-based method for bulk magnetization problems. Although similar to the 2D FFT-based method in many aspects, it uses the magnetic field as the main variable and, unexpectedly, is even easier to implement. Usually, convergence of iterations (18) is very fast, which makes this method efficient.

Our solution of the benchmark problem (magnetization of a cubic superconductor) agrees well with those obtained using more complicated finite element approaches in [19]. For small penetration, our numerical solution for a superconducting ball is close to the known asymptotic solution. An extension of this method to multiply connected geometry was used to model magnetization of a superconductor with a hole, topologically equivalent to torus.

Although in our 3D examples we used only a simple isotropic current-voltage relation, it can be replaced by another one (anisotropic, field dependent, etc.) In particular, to simulate magnetization of stacks and coils the method can be used with the homogenized anisotropic bulk model. Further improvements of this method should, probably, involve a more efficient ODE solver and a more sophisticated approach to problems with multiply connected geometry.

**Acknowledgment.** Discussions with J.I. Vestgården on the 2D FFT-based method are highly appreciated.